\documentclass[superscriptaddress,amsmath, nofootinbib]{revtex4}
\usepackage{amsfonts}
\usepackage{graphicx}
\usepackage[brazil]{babel}
\usepackage[latin1]{inputenc}
\usepackage{amsmath}
\usepackage{amssymb}
\usepackage{lineno}

\begin{document}


\title{Has the problem of the motion of a heavy symmetric top been solved in quadratures?}

\author{Alexei A. Deriglazov }
\email{alexei.deriglazov@ufjf.br} \affiliation{Depto. de Matem\'atica, ICE, Universidade Federal de Juiz de Fora,
MG, Brazil} 


\date{\today}

\begin{abstract}
We have revised the problem of the motion of a heavy symmetric top. When formulating equations of the Lagrange top with the diagonal inertia tensor, the potential energy has more complicated form as compared with that assumed in the literature on dynamics of a rotating body. This implies the corresponding improvements in equations of motion. Using the Liouville's theorem, we solve the improved equations in quadratures and present the explicit expressions for the resulting elliptic integrals. 
\end{abstract}

\maketitle 



\section{Introduction.}
The dynamics of a free rotating body can be formulated starting from an appropriately chosen variational problem for a system of particles subject to kinematic constraints \cite{AAD23, AAD23_1}.  Here we follow the same scheme to deduce equations of a rotating body with fixed point in the gravity field. The analysis is a similar to that presented in \cite{AAD23}, so we only outline it in Sect. II, emphasizing those points that will be important in the subsequent discussion. Being one of the classical problems of non linear dynamics and integrable systems, this issue however is of interest in the modern studies of various aspects related with construction and behavior of spinning particles and rotating bodies in external fields beyond the pole-dipole approximation \cite{Pio_2020,Tib_2019,Car_2018,Sam_2018,Ren_2018,Kim_2023,Cri_2017,Anw_2023,Bis_2022,Mar_2023}. 

Then we concentrate on the case of a symmetrical body (Lagrange top). The Lagrange top is one of the oldest problems of mathematical physics, which always has been considered as the classical example of the system integrable according to Liouville which, moreover, admits a reasonably simple qualitative description on the base of effective potential \cite{Poin,Whit_1917,Mac_1936,Lei_1965,Gol_2000,Grei_2003,Arn_1,Landau_8}. However, when formulating its variational problem with the {\it diagonal inertia tensor}, we observed that potential energy has more complicated form as compared with that assumed in the literature on dynamics of a rigid body. This implies the corresponding improvements in equations of motion.  Because this is a somewhat surprising observation, its validity and comparison with the literature are detailly carried out in Sect. II, by pointing out precisely the place, where an inaccuracy in the analysis was allowed, leading to incorrect equations. 

In Sect. III, using the Liouville's theorem, we solve the improved equations in quadratures and show that the problem can be reduced to calculation of four elliptic integrals. Their explicit form is given in Eqs. (\ref{s18})-(\ref{s21}).

\section{Improved equations of the Lagrange top.}

{\bf Free asymmetric body.} The evolution of a free rotating body in the center-of-mass system can be described by $3+9$ Euler-Poisson equations \cite{AAD23}
\begin{eqnarray}\label{s0}
I\dot{\boldsymbol\Omega}=[I{\boldsymbol\Omega}, {\boldsymbol\Omega}], \label{s0} \qquad    
\dot R_{ij}=-\epsilon_{jkm}\Omega_k R_{im}.  \label{s1}
\end{eqnarray}
The functions $\Omega_i(t)$ are components of angular velocity in the body \cite{Arn_1}, and $R_{ij}(t)$ is orthogonal $3\times 3$\,-matrix\footnote{We use notation from the work \cite{AAD23}. In particular, the scalar product is denoted as 
follows: $({\bf a}, {\bf b})=a_i b_i=a_1 b_1+a_2 b_2+a_3 b_3$. Notation for the vector product: $[{\bf a}, {\bf b}]_i=\epsilon_{ijk}a_j b_k$, 
where $\epsilon_{ijk}$ is Levi-Chivita symbol in three dimensions, with $\epsilon_{123}=+1$.}. Given the solution $R(t)$, the evolution of the body's point ${\bf x}(t)$ is restored according to the rule: ${\bf x}(t)=R(t){\bf x}(0)$, where ${\bf x}(0)$ is the initial position of the point. Due to this, the problem (\ref{s0}) should be solved with the universal initial conditions: $R_{ij}(0)=\delta_{ij}$, ${\boldsymbol\Omega}(0)={\boldsymbol\Omega}_0$. The solutions with other initial conditions are not related to the motions of a rigid body\footnote{Failure to take this circumstance into account leads to a lot of confusion, see \cite{AAD23_3}.}.  Both columns and rows of the matrix $R_{ij}$ have a geometric interpretation. The columns $R(t)=({\bf R}_1(t), {\bf R}_2(t), {\bf R}_3(t))$ form an orthonormal basis rigidly connected to the body. The initial conditions imply that at $t=0$ these columns coincide with the basis vectors ${\bf e}_i$ of the Laboratory system.  The rows, $R^T(t)=({\bf G}_1(t), {\bf G}_2(t), {\bf G}_3(t))$, represent the laboratory basis vectors ${\bf e}_i$ in the body-fixed basis. For example, the functions ${\bf G}_3(t)=(R_{31}(t), R_{32}(t), R_{33}(t))$ are components of ${\bf e}_3$ in the basis ${\bf R}_i(t)$. The Hamiltonian origin of the first-order equations (\ref{s1}) can be established with use of the intermediate formalism \cite{AAD23_2}.

By $I$ in Eq. (\ref{s0}) was denoted the inertia tensor. For the body considered as a system of $n$ particles with coordinates ${\bf x}_N(t)$ and masses $m_N$, $N=1, 2, \ldots , n$,  it is a numeric $3\times 3$\,-matrix defined as follows:
\begin{eqnarray}\label{s2}
I_{ij}\equiv\sum_{N=1}^{n}m_N\left[{\bf x}_N^2(0)\delta^{ij}-x_N^i(0)x_N^j(0)\right]. 
\end{eqnarray}
Generally, $I_{ij}$ is a symmetric matrix 
\begin{eqnarray}\label{s0.02}
I=\left(
\begin{array}{ccc}
I_{11} & I_{12} & I_{13} \\
I_{12} & I_{22} & I_{23} \\
I_{13} & I_{23} & I_{33} 
\end{array}\right),  
\end{eqnarray}
transforming as the second-rank tensor under rotations of the Laboratory system.  So the explicit form of the numeric matrix, that appears in equations (\ref{s0}), depends on the initial position of the body. Equivalently, it can be said that it change when we pass from one Laboratory basis to another one, related by some rotation. Indeed, consider two orthonormal bases related by rotation with help of numeric orthogonal matrix $U^TU=1$: ${\bf e}'_i={\bf e}_kU^T_{ki}$. Coordinates of the body's particles in these bases are related as follows: 
$x'^i=U_{ij}x^j$.
Then Eq. (\ref{s2}) implies that the matrices $I'_{ij}$ and $I_{ij}$, computed in these bases, are related by 
\begin{eqnarray}\label{s0.03}
I'_{ij}=\sum_{N=1}^{n}m_N\left[{\bf x'}_N^2(0)\delta^{ij}-x'^i_N(0)x'^j_N(0)\right]=
U_{ia}\left[\sum_N {\bf x}_N^2(0)\delta^{ab}- m_Nx_N^a(0)x_N^b(0)\right]U_{jb}=U_{ia}I_{ab}U_{jb}, \quad \mbox{or} \quad I'=UIU^T. \quad 
\end{eqnarray}
Adapting the Laboratory system with the position of the body at $t=0$, we can simplify Eqs. (\ref{s0}). Indeed, assume that at the instant $t=0$ the Laboratory axes ${\bf e}_i$ have been chosen in the direction of eigenvectors of the matrix $I_{ij}$. Then the inertia tensor in Eqs. (\ref{s0}) acquires diagonal form 
\begin{eqnarray}\label{s0.01}
I=\left(
\begin{array}{ccc}
I_1 & 0 & 0 \\
0 & I_2 & 0 \\
0 & 0 & I_3 
\end{array}\right). 
\end{eqnarray}
As we saw above, due to initial conditions $R_{ij}(0)=\delta_{ij}$, the axes ${\bf R}_i(t)$ of body-fixed basis at $t=0$ coincide with the Laboratory axes ${\bf e}_i$, and therefore coincide also with the inertia axes.  Since the axes ${\bf R}_i(t)$ and the inertia axes are rigidly connected with the body, they will coincide in all future moments of time.


Let's consider an asymmetric rigid body, that is ($I_1\ne I_2\ne I_3$), and suppose that we describe it using the equations (\ref{s0}), in which the inertia tensor is chosen to be diagonal. This implies that the position of the Laboratory system is completely fixed, as described above. If for some reason we want to choose a different coordinate system, we will forced to use equations (\ref{s0}) with the symmetric matrix (\ref{s0.02}) containing non zero off-diagonal elements instead of diagonal matrix (\ref{s0.01}).

As will be seen further, it is precisely this circumstance that is not taken into account in monographs when formulating the equations of heavy symmetric top and solving them. 

{\bf Heavy symmetric body with a fixed point.} Let's consider a rigid body with one fixed point. It is known \cite{Whit_1917,Mac_1936,Lei_1965,Gol_2000,Poin,Arn_1} that by placing the origin of the Laboratory system at this point, we arrive at the same equations (\ref{s0}) and (\ref{s2}). The only difference is that the inertia tensor must now be calculated with respect to the fixed point.
Further, let gravity act on the body, with the free fall acceleration equal to $a>0$ and directed opposite to the constant unit 
vector ${\bf k}$, see Figure 1(a).  Then the potential energy of the body's particle ${\bf x}_N(t)$ is $am_N({\bf k}, {\bf x}_N(t))$. Summing up the potential energies of the body's points, we get the total energy $b({\bf k}, {\bf z}(t))=b({\bf k}, R(t){\bf z}(0))$. Here $b=aL\mu$, $L$ is the distance from the center of mass to the fixed point, $\mu$ is the total mass of the body and ${\bf z(0)}$ is unit vector in the direction of center of mass at $t=0$. The potential energy give rise to the torque of gravity in the equations of motion, which read now as follows: 
\begin{eqnarray}\label{s4}
I\dot{\boldsymbol\Omega}=[I{\boldsymbol\Omega}, {\boldsymbol\Omega}]+b[R^T{\bf k}, {\bf z}(0)], \label{s4}  \qquad   
\dot R_{ij}=-\epsilon_{jkm}\Omega_k R_{im}.  \label{s5}
\end{eqnarray} 
\begin{figure}[t] \centering
\includegraphics[width=12cm]{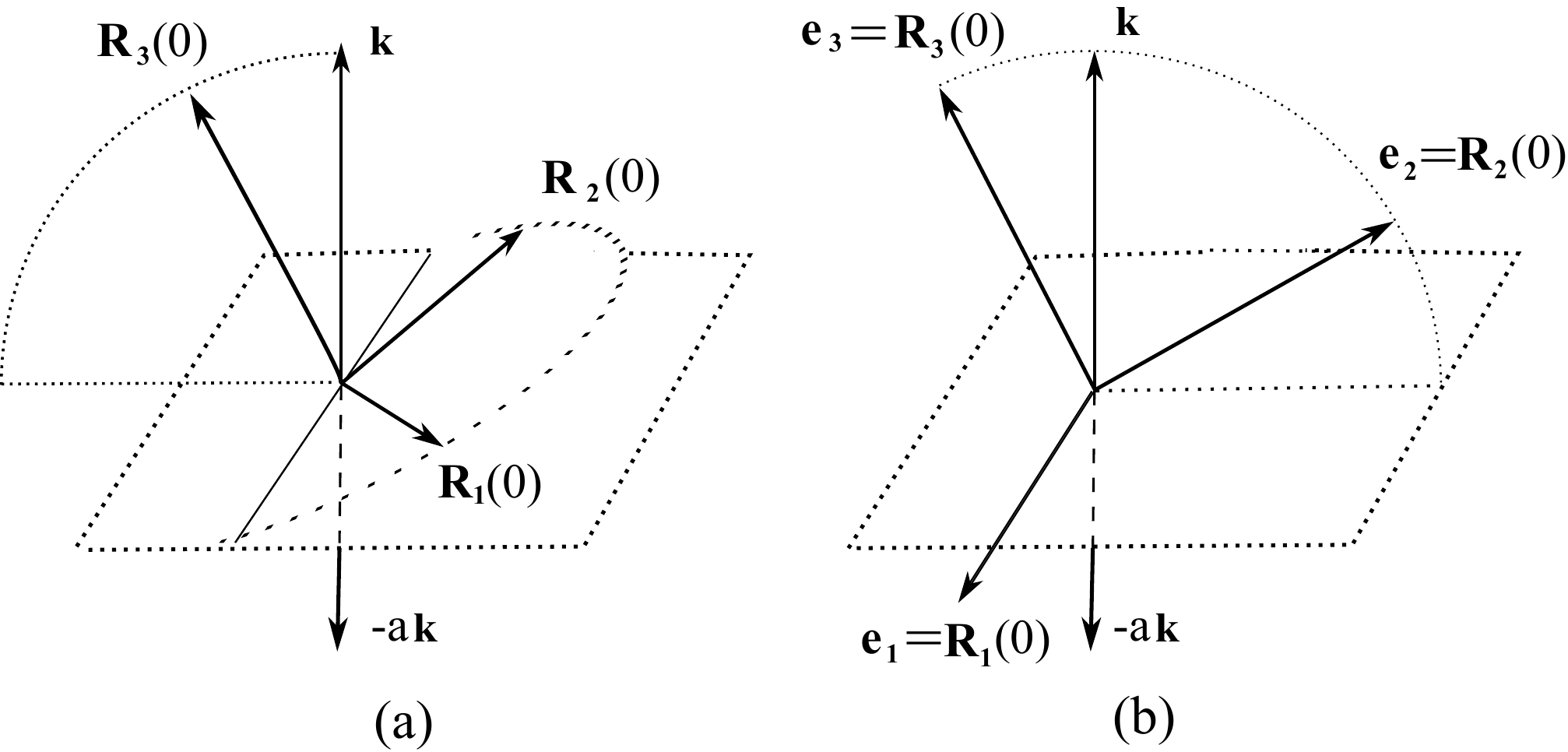}
\caption{(a) - Initial position of a heavy top with orthogonal inertia axes ${\bf R}_1(0), {\bf R}_2(0), {\bf R}_3(0)$. (b) - For the Lagrange top, due to the freedom in the choice of ${\bf R}_1(0)$ and ${\bf R}_2(0)$, the vector ${\bf k}$ can be taken in the form ${\bf k}=(0, k_2, k_3).$}\label{Heavy}
\end{figure}
Let the initial position of the inertia axes ${\bf R}_i(0)$ of the body be as shown in Figure \ref{Heavy}(a). Assuming that the Laboratory axes ${\bf e}_i$ have been chosen in the direction of the inertia axes, the matrix $I$ in Eq. (\ref{s4}) acquires the diagonal form.  Let us denote components of the vector ${\bf k}$ in this basis as $(k_1, k_2, k_3)$. 

If our body is symmetrical, that is $I_1=I_2\ne I_3$, our equations can be further simplified as follows.  Without loss of generality, we can assume  that the vector ${\bf k}$ has the following form: ${\bf k}=(0, k_2, k_3)$. 
Indeed, the eigenvectors and eigenvalues of the inertia tensor $I$ obey the relations $I{\bf R}_i(0)=I_i{\bf R}_i(0)$.  With $I_1=I_2$ we have $I{\bf R}_1(0)=I_2{\bf R}_1(0)$ and $I{\bf R}_2(0)=I_2{\bf R}_2(0)$, then any linear combination $\alpha{\bf R}_1(0)+\beta{\bf R}_2(0)$ also represents an eigenvector with eigenvalue $I_2$. This means that we are free to choose any two orthogonal axes on the plane ${\bf R}_1(0), {\bf R}_2(0)$ as the inertia axes. Hence, in the case $I_1=I_2$ we can rotate the Laboratory axes in the plane $(x^1, x^2)$  without breaking the diagonal form of the inertia tensor. Using this freedom, we can assume that $k_1=0$ for our problem, see Figure 1(b).  

{\bf Lagrange top.} Let's consider the symmetrical body, $I_1=I_2\ne I_3$, and assume that the fixed point was chosen such that center of mass lies on the axis of inertia ${\bf R}_3(t)$. This body is called the Lagrange top \cite{Arn_1}.  This allows further simplify the equations of motion, since by 
construction ${\bf z}(0)=(0, 0, 1)$. 

With these ${\bf k}=(0, k_2, k_3)$ and ${\bf z}(0)=(0, 0, 1)$, the potential energy acquires the form $b(k_2R_{23}+k_3R_{33})$. Using these ${\bf k}$ and ${\bf z}(0)$ in Eqs. (\ref{s4}), we get the final form of equations of motion of a heavy symmetric top for the variables $\Omega_i$ and $R_{ij}$. To compare them with those given in textbooks, we rewrite our equations in terms of Euler angles. To reduce the amount of computation, it is better to work with
the Lagrangian that implies equations (\ref{s4}) as the conditions of its extremum \cite{AAD23}
\begin{eqnarray}\label{hb24}
L= \frac12 I_i(\Omega_i)^2-\frac12 \lambda_{ij}\left[R_{ki}R_{kj}-\delta_{ij}\right]-b(k_2R_{23}+k_3R_{33}), 
\end{eqnarray}
where $\Omega_i\equiv-\frac12 \epsilon_{ijk}(R^T\dot R)_{jk}$.  Let us substitute the expression for $R_{ij}$ in terms of Euler angles \cite{AAD23}
\begin{eqnarray}\label{6.3}
R=\left(
\begin{array}{ccc}
\cos\psi\cos\varphi-\sin\psi\cos\theta\sin\varphi  &  -\sin\psi\cos\varphi-\cos\psi\cos\theta\sin\varphi &\sin\theta\sin\varphi \\
\cos\psi\sin\varphi+\sin\psi\cos\theta\cos\varphi & -\sin\psi\sin\varphi+\cos\psi\cos\theta\cos\varphi & -\sin\theta\cos\varphi \\
\sin\psi\sin\theta & \cos\psi\sin\theta & \cos\theta
\end{array}\right)
\end{eqnarray}
into Eq. (\ref{hb24}).  According to classical mechanics \cite{Arn_1}, this gives an equivalent variational problem. Since the rotation matrix in terms of Euler angles authomatically obeys the  constraint $R^TR={\bf 1}$, the second term of the Lagrangian (\ref{hb24}) vanishes, and we get 
\begin{eqnarray}\label{s7}
L=\frac12 I_2[\dot\theta^2+\dot\varphi^2\sin^2\theta]+ \frac12 I_3[\dot\psi+\dot\varphi\cos\theta]^2-b[k_3\cos\theta-k_2\sin\theta\cos\varphi].
\end{eqnarray}
This Lagrangian, in turn, implies the equations of motion
\begin{eqnarray}
I_3[\dot\psi+\dot\varphi\cos\theta]=m_\psi=\mbox{const},  \label{s6.1} \\ 
\frac{d}{dt}[I_2\dot\varphi\sin^2\theta+m_\psi\cos\theta]+bk_2\sin\theta\sin\varphi=0,   \label{s6.2} \\ 
-I_2\ddot\theta+I_2\dot\varphi^2\sin\theta\cos\theta-m_\psi\sin\theta\dot\varphi+b[k_3\sin\theta+k_2\cos\theta\cos\varphi]=0. \label{s6.3}
\end{eqnarray}

In the textbooks, equations of the Lagrange top follow from a different Lagrangian, the latter does not contain the term  proportional to $k_2$ \cite{Arn_1,Landau_8}
\begin{eqnarray}\label{s7.1}
L_A=\frac12 I_2[\dot\theta^2+\dot\varphi^2\sin^2\theta]+ \frac12 I_3[\dot\psi+\dot\varphi\cos\theta]^2-b\cos\theta. 
\end{eqnarray}
This term is discarded on the base of the following reasoning: to simplify the analysis, choose the Laboratory axis ${\bf e}_3$ in the direction of the vector ${\bf k}$. However, this reasoning does not take into account the presence in the equations of moments of inertia, which have the tensor law of transformation under rotations. Indeed, going back to Eqs. (\ref{s4}), select ${\bf e}_3$ in Figure \ref{Heavy}(b) in the direction of ${\bf k}$, and calculate the components of the inertia tensor. Since the axis of inertia ${\bf R}_3(0)$ does not coincide with ${\bf e}_3$, we obtain a symmetric matrix with non-zero off-diagonal elements (\ref{s0.02}) instead of (\ref{s0.01}). This symmetric matrix should now be used to construct the kinetic part of Lagrangian and hence it appears in the equations of motion. That is, the attempt to simplify the potential energy will lead, instead of (\ref{s7.1}),  to a Lagrangian with a complicated expression for the kinetic energy. 

Does a rotating body have motions that could be described using the equations following from (\ref{s7.1})? The answer is yes: these are solutions with special initial conditions, for which the ${\bf R}_3(t)$\,-axis coincides with ${\bf k}$ at some (finite) instant of time. These are the solutions of an awakened top and its limiting case of a sleeping top, see \cite{AAD23_9}. In the general case, to look for the  solutions that do not pass through ${\bf k}$, one should use the equations following from (\ref{s7}).

Thus, it seems to be necessary to revise the problem of the motion of a Lagrange top and correct this drawback.

\section{Integrability in quadratures according to Liouville.}
Eq. (\ref{s6.1}) states that $m_\psi$ is the integral of motion of the theory (\ref{s7}). Besides, the variable $\psi(t)$ does not enter into the remaining equations (\ref{s6.2}) and (\ref{s6.3}). So we can look for the effective Lagrangian that implies the equations (\ref{s6.2}) and (\ref{s6.3}) for the variables $\theta$ and $\varphi$. This is as follows: 
\begin{eqnarray}\label{s7.2}
L_2=\frac12 I_2[\dot\theta^2+\dot\varphi^2\sin^2\theta]+m_\psi\dot\varphi\cos\theta -b[k_3\cos\theta-k_2\sin\theta\cos\varphi], 
\end{eqnarray}
where $m_\psi$ is considered as a constant.  To apply the Liouville's theorem, we need to rewrite the effective theory in the Hamiltonian formalism. Introducing the conjugate momenta $p_\theta=\partial L/\partial\dot\theta$ and $p_\varphi=\partial L/\partial\dot\varphi$ for the configuration-space variables $\theta$ and $\varphi$, we get
\begin{eqnarray}\label{s8}
p_\theta=I_2\dot\theta, \quad \Rightarrow \quad \dot\theta=\frac{1}{I_2}p_\theta; \qquad 
p_\varphi=I_2\dot\varphi\sin^2\theta+m_\psi\cos\theta, \quad \Rightarrow \quad \dot\varphi=\frac{p_\varphi-m_\psi\cos\theta}{I_2\sin^2\theta}.
\end{eqnarray}
The Hamiltonian of the system is constructed according the standard rule: $H=p\dot q-L(q, \dot q)$. Its explicit form is as follows:
\begin{eqnarray}\label{s9}
H=\frac{1}{2I_2}[p^2_\theta+\tilde p_\varphi^2]+bU(\theta, \varphi), 
\end{eqnarray}
where it was denoted
\begin{eqnarray}\label{s10}
\tilde p_\varphi\equiv\frac{p_\varphi-m_\psi\cos\theta}{I_2\sin\theta}, \qquad U(\theta, \varphi)\equiv k_3\cos\theta-k_2\sin\theta\cos\varphi.
\end{eqnarray}
The Hamiltonian equations can be obtained now according the standard rule: $\dot z=\{z, H\}$, where $z$ is any one of the phase-space variables, and the nonvanishing Poisson brackets are 
$\{\theta, p_\theta\}=1$, $\{\varphi, p_\varphi\}=1$.

The effective theory admits two integrals of motion. They are the energy 
\begin{eqnarray}\label{s12}
E=\frac{1}{2I_2}[p^2_\theta+\tilde p_\varphi^2]+bU(\theta, \varphi), 
\end{eqnarray}
and the projection of angular momentum on the direction of vector ${\bf k}$
\begin{eqnarray}\label{s13}
m_\varphi=k_2p_\theta\sin\varphi-U'_\theta\tilde p_\varphi+m_\psi U, \quad \mbox{where} \quad U'_\theta\equiv\frac{\partial U}{\partial\theta}. 
\end{eqnarray}
The preservation in time of $m_\varphi$ can be verified by direct calculation\footnote{In terms of the variables $\Omega_i$ and $R_{ij}$ the integral of motion $m_\varphi$ reads as follows: $m_\varphi=({\bf k}, RI{\boldsymbol\Omega})\equiv({\bf k}, {\bf m})$, where ${\bf m}=RI{\boldsymbol\Omega}$ is the angular momentum of the top. Its preservation in time $\dot m_\varphi=0$ follows also from Eqs. (\ref{s4}).}. 
The remarcable property of the integrals of motion is that they have vanishing Poisson bracket, $\{ E, m_\varphi\}=0$, as it can be confirmed by direct calculation. Accoding to the Liouville's theorem \cite{Arn_1,Fom_2004}, this implies that a general solution to the equations of motion can be found in quadratures (that is calculating integrals of some known functions and doing the algebraic operations). Our aim now will be to present the explicit form of the integrals in question. 

According to the algorithm used in the proof of Liouville's theorem \cite{Fom_2004}, we need to resolve Eqs. (\ref{s12}), (\ref{s13}) with respect to $p_\theta$ and $p_\varphi$. Using the identities    
\begin{eqnarray}\label{s14}
(U'_\theta)^2=k_3^2+k_2^2\cos^2\varphi-U^2, \quad (U'_\theta)^2+k_2^2\sin^2\varphi=1-U^2, \quad U''_{\theta\theta}=-U,
\end{eqnarray}
we get the solution 
\begin{eqnarray}\label{s15}
p_\varphi=m_\psi\cos\theta+\sin\theta\frac{-U'_\theta(m_\varphi-m_\psi U)+
\epsilon k_2\sin\varphi\sqrt{2I_2(1-U^2)(E-bU)-(m_\varphi-m_\psi U)^2}}{1-U^2}, \cr
p_\theta=\frac{k_2\sin\varphi(m_\varphi-m_\psi U)+
\epsilon U'_\theta\sqrt{2I_2(1-U^2)(E-bU)-(m_\varphi-m_\psi U)^2}}{1-U^2},  \qquad \qquad \qquad \quad
\end{eqnarray}
where $\epsilon=\pm 1$. 
Next we need to integrate these functions along any curve connecting the origin of configuration space with a point $(\theta, \varphi)$. Taking the curve to be a pair of intervals, $(0, 0)\rightarrow(\theta, 0)\rightarrow(\theta, \varphi)$, we obtain the following function
\begin{eqnarray}\label{s16}
\Phi(\theta, \varphi, E, m_\varphi)=\int_{0}^{\theta}p_\theta(\theta', 0, E, p_\varphi)d\theta' +
\int_{0}^{\varphi}p_\varphi(\theta, \varphi', E, p_\varphi)d\varphi'.
\end{eqnarray}
Then the general solution to the equations of motion of the theory (\ref{s7.2}) is obtained resolving Eqs. (\ref{s15}) together with  the equations 
$c_1=-t+\partial\Phi/\partial E$, $c_2=\partial\Phi/\partial m_\varphi$ 
with respect to the phase-space variables $\theta, p_\theta, \varphi, p_\varphi$. Using (\ref{s16}), the last two equations read as follows: 
\begin{eqnarray}\label{s17}
c_1=-t+\int_{0}^{\theta}\partial_Ep_\theta(\theta', 0, E, m_\varphi)d\theta' +
\int_{0}^{\varphi}\partial_Ep_\varphi(\theta, \varphi', E, m_\varphi)d\varphi', \cr
c_2=\int_{0}^{\theta}\partial_{m_\varphi}p_\theta(\theta', 0, E, m_\varphi)d\theta' +
\int_{0}^{\varphi}\partial_{m_\varphi}p_\varphi(\theta, \varphi', E, m_\varphi)d\varphi'. 
\end{eqnarray}
Calculating the partial derivatives indicated in these integrals, we get
\begin{eqnarray}\label{s18}
\int_{0}^{\theta}\partial_Ep_\theta(\theta', 0)d\theta'=\epsilon I_2\int\frac{dU}{\sqrt{2I_2(1-U^2)(E-bU)-(m_\varphi-m_\psi U)^2}},
\end{eqnarray}
\begin{eqnarray}\label{s19}
\int_{0}^{\theta}\partial_{m_\varphi}p_\theta(\theta', 0)d\theta'=
-\epsilon\int\frac{(m_\varphi-m_\psi U)dU}{(1-U^2)\sqrt{2I_2(1-U^2)(E-bU)-(m_\varphi-m_\psi U)^2}},
\end{eqnarray}
\begin{eqnarray}\label{s20}
\int_{0}^{\varphi}\partial_Ep_\varphi(\theta, \varphi')d\varphi'=
-\epsilon I_2k_2\sin\theta\int\frac{d(\cos\varphi')}{\sqrt{2I_2(1-U^2)(E-bU)-(m_\varphi-m_\psi U)^2}},
\end{eqnarray}
\begin{eqnarray}\label{s21}
\int_{0}^{\varphi}\partial_{m_\varphi}p_\varphi(\theta, \varphi')d\varphi'=
-\sin\theta\int_{0}^{\varphi}\frac{U'_\theta d\varphi'}{1-U^2}+
\epsilon k_2\sin\theta\int\frac{(m_\varphi-m_\psi U)d(\cos\varphi')}{(1-U^2)\sqrt{2I_2(1-U^2)(E-bU)-(m_\varphi-m_\psi U)^2}}.
\end{eqnarray}
These are elliptic integrals over the indicated integration variables.

\section{Conclusion.} 
When formulating and solving the equations of motion of a rotating body with the inertia tensor chosen in the diagonal form, one should keep in mind the tensor law of transformation of the moments of inertia $I_1, I_2, I_3$ under rotations. We observed that for the Lagrange top this leads to the potential energy that depends on the Euler angles $\varphi$ and $\theta$. The potential energy (see the last term in (\ref{s7})) is different from that assumed in textbooks (see the last term in (\ref{s7.1})). 
As far as I know, this drawback has not yet been noticed and corrected in the literature. So we revised the problem of the motion of a Lagrange  top and corrected this drawback. We confirmed that the improved equations of motion
(\ref{s6.1})-(\ref{s6.3}) are integrable according to Liouville, and reduced the problem to the elliptic integrals (\ref{s18})-(\ref{s21}). 

Probably for the first time in the monographic literature the equations following from the Lagrangian (\ref{s7.1})  were discussed in details by MacMillan in \cite{Mac_1936}. In the absence of analytical solution in elementary functions, MacMillan performed analysis of integrals of motion and of effective potential, reducing the problem to the study of a polinomial of degree 3. The results of this qualitative analysis are summarized in Figs. 60-62 of his book, and then reproduced in many other textbooks \cite{Gol_2000, Grei_2003, Arn_1, Landau_8}. In this respect we point out that the integrals of motion (\ref{s12}) and (\ref{s13}), which follow from the improved Lagrangian (\ref{s7}), imply the analysis of a polinomial of degree 6.

\begin{acknowledgments}
The work has been supported by the Brazilian foundation CNPq (Conselho Nacional de
Desenvolvimento Cient\'ifico e Tecnol\'ogico - Brasil). 
\end{acknowledgments}


\begin{thebibliography}{99} 

\bibitem{AAD23} A. A. Deriglazov, {\it Lagrangian and Hamiltonian formulations of asymmetric rigid body, considered as a constrained system}, Eur. J. Phys. https://doi.org/10.1088/1361-6404/ace80d; arXiv:2301.10741.   

\bibitem{AAD23_1} A. A. Deriglazov, {\it Geodesic motion on the symplectic leaf of SO(3) with distorted e(3) algebra and Liouville integrability of a free rigid body},  Eur. Phys. J. C (2023) 83:265; arXiv:2302.04828. 

\bibitem{Pio_2020} P. Kosi\'nski and P.  Ma\'slanka, {\it Relativistic symmetries and Hamiltonian formalism},  Symmetry {\bf 12} (2020) 1810.  

\bibitem{Tib_2019} Tiberiu Harko, Shi-Dong Liang, {\it  Energy-dependent noncommutative quantum mechanics}, 
Eur. Phys. J. {\bf C 79} (2019) 4, 300;   arXiv:1903.06776. 

\bibitem{Car_2018} C. Villalpando, S. K. Modak, {\it Minimal length effect on the broadening of free wave-packets and its physical implications}, 
Phys. Rev. {\bf D} 100 (2019) 2, 024054;  arXiv:1812.06112. 

\bibitem{Sam_2018} S. Kov\'acik, P. Presnajder, {\it  Alternative description of magnetic monopoles in quantum mechanics},  Eur. Phys. J. {\bf C 78} (2018) 9, 745;  arXiv:1804.04015.  

\bibitem{Ren_2018} Ya-Jie Ren, Kai Ma, {\it Influences of the coordinate dependent noncommutative space on charged and spin currents},  Int. J. Mod. Phys. {\bf A 33} (2018) 16, 1850093;  arXiv:1802.10452. 

\bibitem{Kim_2023} Joonhwi Kim, {\it  An ambitwistor for Kerr I: Zig-Zag symplectic perturbation theory},   arXiv:2301.06203. 

\bibitem{Cri_2017}   Christoph K\"ohn, {\it The Planck length and the constancy of the speed of light in five dimensional space parametrized with two time coordinates},  JHEP Grav. Cosmol. 3 (2017) 635-650;  arXiv:1612.01832. 

\bibitem{Anw_2023} A. Chakraborty, {\it Emergent geometric phase in time-dependent noncommutative quantum system},  arXiv:2306.08467. 

\bibitem{Bis_2022} B. Chakraborty, P. Nandi, S. Kumar Pal, A. Chakraborty, {\it Our trysts with "Bal" and noncommutative geometry}, arXiv:2212.06548.

\bibitem{Mar_2023} M. Marrocco, {\it "A call to action": Schr\"odinger's representation of quantum mechanics via Hamilton's principle}, American Journal of Physics, {\bf 91} (2023) 2, 110-115.  

\bibitem{Poin} L. Poinsot,  {\it Theorie Nouvelle de la Rotation des Corps},  (Bachelier, Paris, 1834); English 
translation: https://hdl.handle.net/2027/coo.31924021260447.  

\bibitem{Whit_1917} E. T. Whittaker, {\it A treatise on the analytical dynamics of particles and rigid bodies}, (Cambridge: at the University press, 1917). 

\bibitem{Mac_1936} W. D. MacMillan, {\it Dynamics of rigid bodies}, (Dover Publications Inc., New-York, 1936). 

\bibitem{Lei_1965} E. Leimanis, {\it The general problem of the motion of coupled rigid bodies about a fixed point}, (Springer-Verlag, 1965). 

\bibitem{Gol_2000} H. Goldstein, C. Poole and J. Safko, {\it Classical mechanics}, Third edition, (Addison Wesley, 2000) 

\bibitem{Grei_2003} W. Greiner, {\it Classical mechanics}, (Springer-Verlag New York Inc. 2003).  

\bibitem{Arn_1} V. I. Arnold, \textit{Mathematical methods of classical mechanics},
2nd edn. (Springer, New York, NY, 1989). 

\bibitem{Landau_8} L. D. Landau and E. M. Lifshitz, {\it Mechanics}, Volume 1, third edition, (Elsevier, 1976).

\bibitem{AAD23_2} A. A. Deriglazov, {\it Poincar\'e-Chetaev equations in the Dirac's formalism of constrained systems}, arXiv:2302.12423. 

\bibitem{Fom_2004} A. V. Bolsinov and A. T. Fomenko, {\it Integrable Hamiltonian systems}, (Charman and Hall/CRC, 2004).

\bibitem{AAD23_3} A. A. Deriglazov, {\it Comment on the Letter "Geometric Origin of the Tennis Racket Effect'' by P. Mardesic, et al, Phys. Rev. Lett. 125, 064301 (2020)}, arXiv:2302.04190. 

\bibitem{AAD23_9} A. A. Deriglazov, {\it Lagrange top: integrability according to Liouville and examples of analytic solutions}, arXiv:2306.02394.

\end{thebibliography}
\end{document}